\documentstyle[psfig,twocolumn,aps]{revtex}
\begin{document}
\preprint{DNA Transport/Bader et al.}
\title{DNA Transport by a Micromachined Brownian Ratchet Device}

\author{Joel~S.~Bader$^{*,\dagger}$, Richard~W.~Hammond,$^\dagger$
Steven~A.~Henck,$^\dagger$ Michael~W.~Deem,$^{\dagger,\ddagger}$
Gregory~A.~McDermott,$^\dagger$ James~M.~Bustillo,$^\S$
John~W.~Simpson,$^\dagger$ Gregory~T.~Mulhern,$^\dagger$
Jonathan~M.~Rothberg$^\dagger$}

\date{\today}
\maketitle

\noindent $^*$Author to whom correspondence should be addressed.

\noindent $^\dagger$CuraGen Corporation, 555 Long Wharf Drive, New
Haven, CT\ \ 06511.

\noindent $^\ddagger$Present address: Department of Chemical
Engineering, University of California, Los Angeles, CA\ \ 90095.

\noindent $^\S$Department of Electrical Engineering and Computer
Science, University of California, Berkeley, CA\ \ 94720.

\noindent Classification

Biological Sciences: Biophysics

Physical Sciences: Chemistry

\noindent Corresponding Author

Joel S. Bader, CuraGen, 555 Long Wharf Drive, New Haven, CT,
06511.

Tel. (203)401-3330x236; Fax (203)401-3351; Email
jsbader@curagen.com

\noindent Manuscript information: 15 text pages, 4 figures, no
tables.

\noindent Counts: 92 words in abstract; 27,800 characters in paper
(counting spaces); 35,720 characters counting figure and equation
requirements.

\noindent Abbreviations: all standard.

\clearpage

\begin{abstract}
We have micromachined a silicon-chip device that transports DNA
with a Brownian ratchet that rectifies the Brownian motion of
microscopic particles.  Transport properties for a DNA 50mer agree
with theoretical predictions, and the DNA diffusion constant
agrees with previous experiments. This type of micromachine could
provide a generic pump or separation component for DNA or other
charged species as part of a microscale lab-on-a-chip. A device
with reduced feature size could produce a size-based separation of
DNA molecules, with applications including the detection of single
nucleotide polymorphisms.
\end{abstract}

\clearpage

\narrowtext

\section{Introduction}

The Human Genome Project aims to provide the complete sequence of
the 3 billion base-pairs of the human genome.  While the dominant
method for analyzing DNA fragments remains gel electrophoresis,
new technologies that have the potential to increase the rate and
decrease the cost of DNA sequencing and analysis, such as mass
spectrometry and hybridization arrays, are critical to the
project's success \cite{rmh97}.

Here we describe a novel method of DNA transport and separation
based on a Brownian ratchet.  As described originally by
Smoluchowski and noted by Feynman, a Brownian particle can undergo
net transport on a potential energy surface that is externally
driven to fluctuate between several distinct states
\cite{smol,feynman}. Brownian ratchets have attracted
theoretical attention \cite{ba96,ab93,ab96,a97,m93,dhr94,hb96,pcpa94,jap97} due to their
description of molecular motors \cite{hhv89,ks93,sssb93}
and to their similarity with phenomena termed stochastic resonance
and resonance activation
\cite{ghj98,dg92}.

Brownian ratchets have been demonstrated to transport  $\mu$m- to
mm-sized particles using dielectrophoresis \cite{rsap94}, optical
tweezers \cite{fbkl95}, and electrocapillary forces \cite{gis96}
to generate ratchets. Other devices based on entropic ratchets
\cite{sgn97} or physical barriers \cite{e98,da98}
have been proposed as well.  More recently, a
geometrical sieve device has been used to separate
phospholipids\cite{boxer99}.

Despite these successes, the Brownian ratchet mechanism has not
before proved capable of transporting DNA fragments in the size
ranges applicable to DNA sequencing ($<$ 1000 nt) because the
interactions used to establish the ratchet potential were too
weak.  In contrast to previous devices using polarization
interactions to generate ratchets \cite{rsap94,fbkl95}, we have
fabricated a device that uses charge-charge interactions to
generate the ratchet potential. As seen below, the charge-charge
interactions have sufficient strength to establish ratchets that
can trap small DNA fragments.

The ratchet-like wells that trap DNA are generated by charging a
series of patterned electrodes. When the electrodes are
discharged, the traps vanish and the molecules undergo Brownian
motion. Next the potential is re-applied, and the particles again
collect in the traps. A spatial asymmetry in the shape of each
ratchet-shaped well rectifies the Brownian motion and produces net
transport as the on-off cycle is repeated.  Each molecule's
transport rate depends on its diffusion constant, allowing the
possibility of size-based separations.  We have implemented the
device by microfabrication on a silicon chip.

This report describes the Brownian ratchet theory, provides a
derivation of the transport rate, and presents experimental
results for a single sized DNA oligomer.  Greater details
regarding the fabrication methods and a more extensive
presentation of the experimental results for a variety devices and
oligomer sizes are available elsewhere \cite{electrophoresis}.

\section{Theory and Methods}

\begin{figure}[htbp]
\centering
\leavevmode
\psfig{file=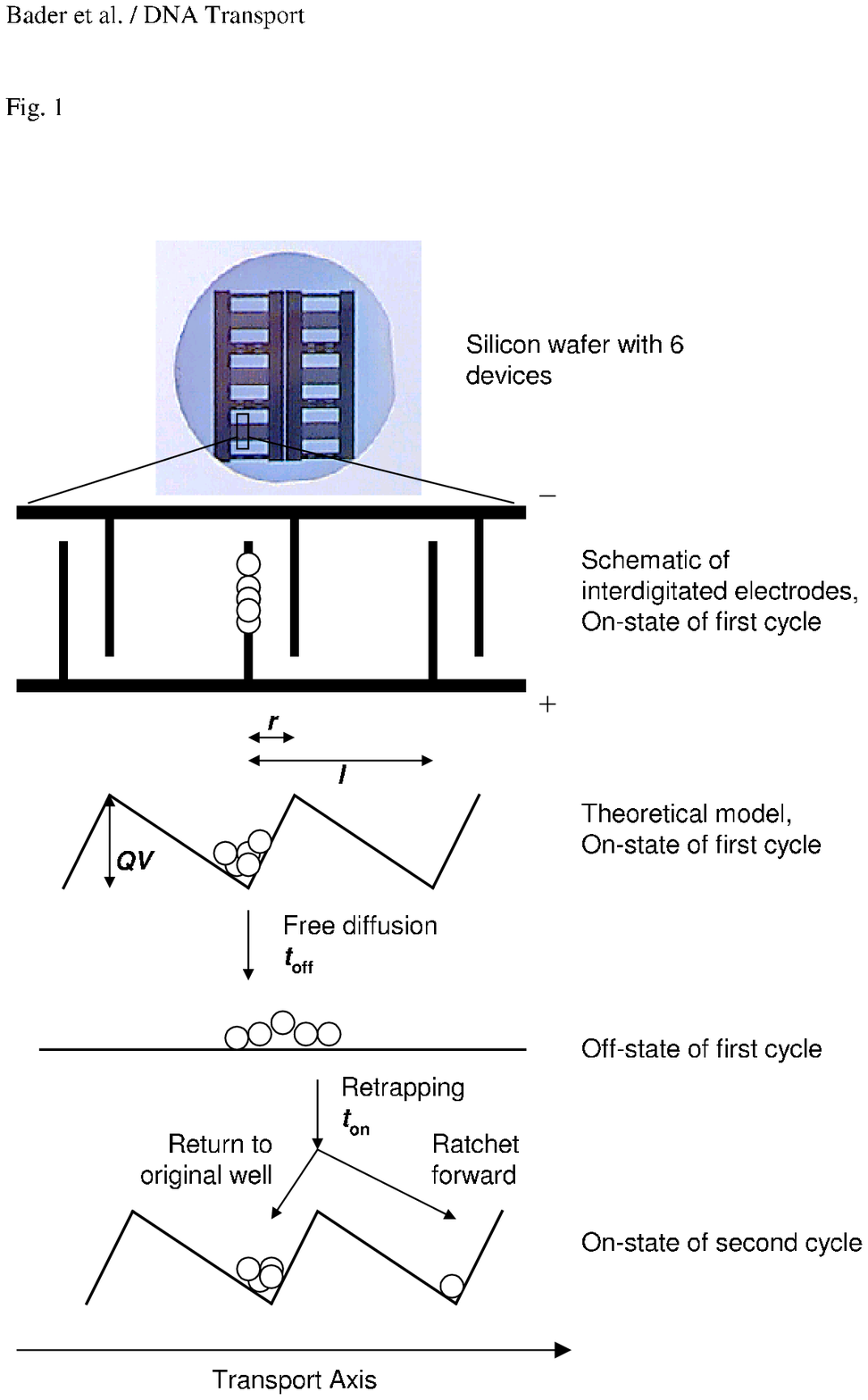,width=3in}
\caption{The Brownian ratchet device is shown in schematic.  Modulating
the electric potential at the electrodes generates a ratchet-like potential
energy surface for charged molecules like DNA.  Cycling the ratchet between
an on-state and an off-state generates transport.}
\label{f:device}
\end{figure}

A silicon wafer with six micromachined devices is shown in
Fig.~\ref{f:device}, with a schematic design below.  The electric
potential that creates the ratcheting traps is generated by two
arrays of interdigitated electrodes that are perpendicular to the
transport axis.  The two sets of arrays each extend from bond pads
on opposite sides of the device.  The spacing between electrodes
extending from the same bond pad is $l$.  The asymmetric pattern
creates two different spacings, $r$ and $l-r$, between electrodes
extending from opposite bond pads.  The smaller spacing $r$
defines the feature size of the device.

As shown in also shown in Fig.~\ref{f:device}, a simplified
one-dimensional description of the potential approximates the
electrodes as infinitely thin wires. To simplify the theoretical
model, we have neglected the finite width of the electrode, the
dependence of the electric potential on the distance normal to the
surface, and the corresponding detailed calculation of the
potential along the transport axis.

When a voltage difference $V$ is applied across the two electrode
arrays during the on-state, with duration $t_{\rm on}$,
sawtooth-shaped ratcheting traps are created for charged
particles.  The electrodes are discharged to $V = 0$ during the
off-state, with duration $t_{\rm off}$, and particles undergo
isotropic Brownian diffusion.  When the potential is re-applied,
the particles are again trapped in potential wells.  The times
$t_{\rm on}$ and $t_{\rm off}$ are within the low-frequency,
quasi-static approximation; nonadiabatic effects and current
reversals, reported elsewhere \cite{bhk94,jkh96}, are not
applicable in this regime. Due to the asymmetry of the sawtooth
shape and the choice of $t_{\rm off}$, a particle starting from
well 0 has a measurable probability to be trapped in well~1 and
virtually zero probability to be trapped in well~$-1$. Transport
can be generated by repetitive cycling between the on-state and
the off-state. The transport properties of a particle are
determined by its diffusion constant $D$ and its charge $Q$, along
with the thermal energy $k_B T$ and the device parameters defined
previously.

Each cycle begins by applying the potential $V$ for a time $t_{\rm
on}$ that is sufficient to localize particles at the bottoms of the
trapping wells.  A particle at the barrier-top drifts down the side of
the sawtooth of length $l-r$ to find the bottom of the well, which
defines the time required for complete trapping,
\begin{equation}
t_{\rm on} = (l-r)^2 k_B T / Q V D \ ,
\label{e:t_on}
\end{equation}
according to overdamped Brownian motion.  At the end of the
trapping, the particle distribution at the bottom of a well is
assumed to be much narrower than the feature size $r$.

In the next phase of the cycle, the potential is turned off for
time $t_{\rm off}$.  When the potential is re-applied, particles
that have diffused further than the barrier to the left (roughly
distance $l$ away) will hop to the previous well, and particles
that have diffused further than the barrier to the right (roughly
distance $r$ away) will be transported to the next well.  Since $r
\ll l$, we can choose an intermediate time $t_{\rm off}$ such that
$t_{\rm off} ~ r^2/2D \ll l^2/2D$ and particles always move right,
never left.

To compute the probability $\alpha$ that a particle moves one well
to the right in a single cycle, we note that the effective
distance $r_{\rm eff}$ it must travel is between $r$ (the inner
edge-to-edge distance between electrodes) and $3r$ (the outer
edge-to-edge distance).  The probability distribution (or
equivalently the Greens function) for a particle starting at the
origin undergoing one-dimensional Brownian motion for time $t_{\rm
off}$ is
\begin{equation}
P(x; t_{\rm off}) = \frac{ \exp[-x^2 / 4 D t _{\rm off}]}{ \sqrt{4
\pi D t_{\rm off} } }.
\end{equation}
An expression for $\alpha$ is then obtained by integrating $P(x;
t_{\rm off})$ from $r_{\rm eff}$ to infinity:
\begin{equation}
\alpha = \frac{1}{2} {\rm erfc}( \sqrt{ r_{\rm eff}^2 / 4 D t_{\rm
off}} ) = \frac{1}{2} {\rm erfc} ( \sqrt{t_r / 2 t_{\rm off}} ).
\label{e:alpha}
\end{equation}
The distance $r_{\rm eff}^2$ has been written in terms of the
characteristic diffusion time $2 D t_r$, where $t_r = r_{\rm
eff}^2/2D$ is the time required to diffuse a distance equal to the
short side $r_{\rm eff}$ of the trapping well.  In this
derivation, we assumed that $t_{\rm off}$ is short enough that
particles diffuse less than a single ratchet, i.e. $P(x; t_{\rm
off}) \approx 0$ for $x > l$.  An expression for $\alpha$ valid in
the limit of large $t_{\rm off}$ is $\alpha = \sum_{k =
-\infty}^{\infty} k \int_{(k-1)l + r_{\rm eff}}^{kl + r_{\rm eff}}
dx\, P(x; t_{\rm off})$.

After each cycle, the particle distribution shifts to the right the
distance $\alpha l$.  After $n$ cycles, the envelope of the
distribution of particles in each well evolves as a Gaussian with
center $x(n)$ and square width $\sigma^2(n)$:
\begin{mathletters}
\begin{equation}
x(n) = n l \alpha, \label{e:center} \end{equation}
\begin{equation} \sigma^2(n) = n l^2 \alpha (1 - \alpha).
  \label{e:width} \end{equation}
\end{mathletters}
Here we have assumed that all the particles are in well 0 at the start
of the first cycle.

\begin{figure}[htbp]
\centering
\leavevmode
\psfig{file=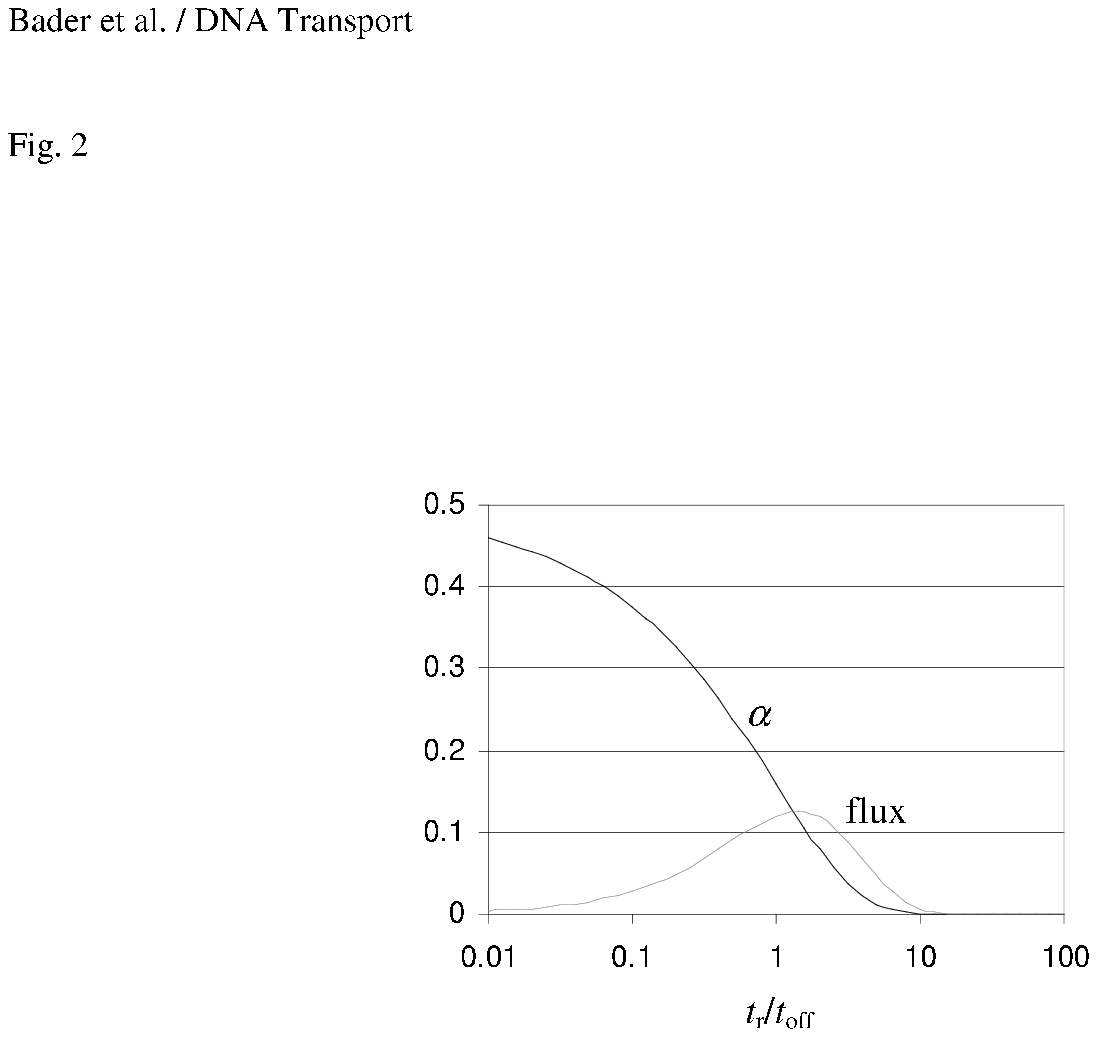,width=3in}
\caption{The probability $\alpha$ that a particle hops one well to
the right during a single cycle of device operation is shown as a
function of the ratio $t_r/t_{\rm off}$ (black line).  For large
$t_{\rm off}$, $\alpha$ approaches 0.5 as we do not consider
back-diffusion.  The flux in units of the characteristic time
$t_r$ is $\alpha t_r/(t_{\rm on} + t_{\rm off})$ and is shown for
$t_{\rm on} = t_{\rm off}/3$ (grey line).  The flux is reminiscent
of a stochastic resonance with a maximum when $t_{\rm off} \approx
t_r$.} \label{f:flux}
\end{figure}

Both $\alpha$ and the steady-state flux of particles through the
device,
\begin{equation}
{\rm flux} = \frac{\alpha}{t_{\rm on} + t_{\rm off} },
\label{e:flux}
\end{equation}
are plotted in Fig.~\ref{f:flux}. The transport fraction $\alpha$
(black line) increases with $t_{\rm off}$ and approaches a maximum
value of 1/2 when back-diffusion is neglected.  The flux in units
of the characteristic time $t_r$ is $\alpha t_r/(t_{\rm on} +
t_{\rm off})$ and is shown for $t_{\rm on} = t_{\rm off}/3$.  The
flux is non-monotonic and exhibits a maximum when $t_{\rm off}
\approx t_r$. Other voltage modulations more complicated than the
on-off pattern described here are also possible and can change the
direction of particle flow according to particle size \cite{ba96}.
Other similar types of non-monotonic behavior have been termed
stochastic resonance and resonance
activation\cite{ghj98,dg92} although they may
also be described by dispersion and linear
response\cite{dispersion}.


Devices were fabricated from Pt, a non-reactive,
corrosion-resistant metal chosen to avoid electrolysis of water,
using relatively standard micromachining technologies
\cite{electrophoresis}. The fabrication began with thermally
oxidized 100~mm diameter silicon wafers.  A 200~\AA\ thick Ti
layer was used as an adhesion layer between the subsequent Pt
layer and the silicon dioxide. Next, a 200~nm thick layer of Pt
was deposited on top of the Ti. The electrodes were defined in the
metal layers using photolithography and ion milling.  For the
devices used in this work, the electrodes and the gaps between
nearest electrodes were 2~$\mu$m and the spatial period was
20$\mu$m.

Oligomers labeled with fluorescent rhodamine dye (Amitof Biotech
Inc., Boston, MA) were placed on the surface of the chip at 4
pmole/$\mu$l in deionized water.  A microscope slide cover glass
(Macalaster Bicknell, New Haven, CT) cut to size was used to
confine the solution to a uniform thickness of approximately
10~$\mu$m.  A sealing compound was used to obtain a liquid tight
seal along the edges, leaving the ends open.  Square wave
modulation (Synthesized Function Generator DS345, Stanford
Research Systems, Sunnyvale, CA) with an amplitude of 1.6~V and
offset of 0.8~V (one set of electrodes at 0~V and the second set
at 1.6~V) was applied to the electrodes to generate the flashing
ratchet potential. Frequencies ranging from 0.7~Hz to 8~Hz were
used, with a ratio $t_{\rm on}/t_{\rm off} = 1/3$.

Video images of the analyte fluorescence were used to record the
DNA transport.  Images were captured using a low light imaging CCD
camera (MTI VE1000SIT) mounted on an epi fluorescence microscope
(Zeiss Axioskop, Germany) using a 10$\times$ Fluor objective.  The
chip was illuminated using the output of a 50~W mercury lamp
filtered with a green band pass filter.  The brightfield
fluorescence was imaged through a red low pass filter. The video
images were recorded on video tape and transferred to a PC using a
composite color PCI bus frame grabber (DT3153 Data Translation,
Inc., Marlboro, MA).  The fluorescence intensity resulting from
the fluorescently labeled DNA fragments was analyzed in a line
across the video image synchronized to the $t_{\rm on}$ period (HL
Image++97, Western Vision Software, Utah).

According to Eq.~\ref{e:t_on}, the expected time required to focus
a DNA 50mer is 0.006~sec, based on a thermal energy of 26~meV, a
charge of 1$|e|^-$ per nucleotide yielding $QV =$ 80~eV, and a
diffusion constant of $1.8 \times 10^{-7}{\rm cm}^2/{\rm sec}$
(estimated from a rhodamine-labeled 30mer on a quartz surface
\cite{xy97}). We visually ascertained that the trapping time
$t_{\rm on}$ was sufficient to permit complete focusing of the DNA
on the positive electrodes, even for the fastest switching rate of
8~Hz  ($t_{\rm on} =$ 0.03~sec).

\section{Results and Discussion}

\begin{figure}[htbp]
\centering
\leavevmode
\psfig{file=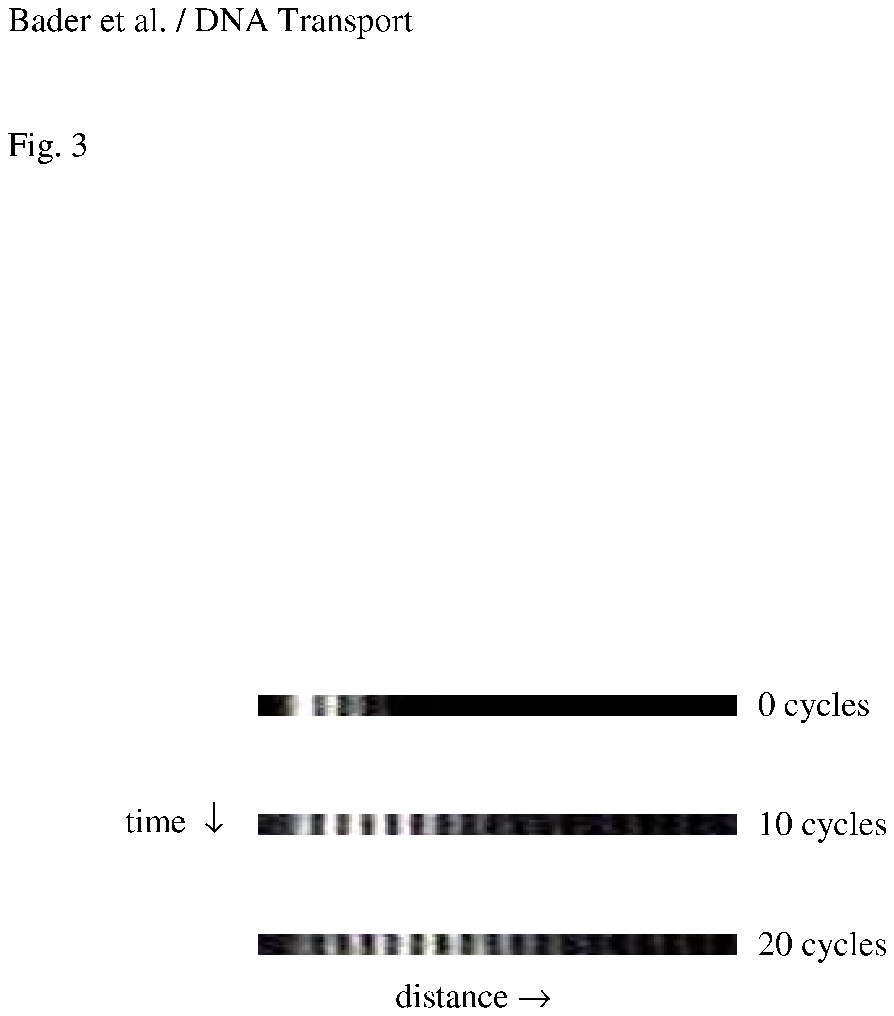,width=3in}
\caption{ Three images are shown from a typical experiment using a
device with 2~$\mu$m electrodes and a 0.7~Hz switching frequency
to transport a rhodamine-labeled DNA 50mer. These images were
saved during the trapping phase of the cycle, and fluorescence
maxima are clearly seen from DNA molecules captured on the
positive electrodes. At the start of the experiment, the DNA
oligomers are focused on left-most three electrodes.  As the
potential cycles between on and off states, the packet moves to
the right and broadens.} \label{f:frames}
\end{figure}

In Fig.~\ref{f:frames} we show three images from a typical
experiment using a device with 2~$\mu$m electrodes and a 0.7~Hz
switching frequency to transport a rhodamine-labeled DNA 50mer.
These images were saved during the trapping phase of the cycle,
and fluorescence maxima are clearly seen from DNA molecules
captured on the positive electrodes. At the start of the
experiment, the DNA oligomers are focused on left-most three
electrodes.  As the potential cycles between on and off states,
the packet moves to the right and broadens.  The transport rate
$\alpha$ can be estimated by noting that the intensity maximum
moves from electrode 1 to electrode 3 after 10 cycles, then to
electrode 5 after 10 more cycles, yielding $\alpha \approx 0.2$.
For a more precise value, we extracted the intensity profile
across the image, set a baseline at the 85$^{\rm th}$ percentile
of intensity, calculated the average position $x(n)$ from the
intensity above baseline, normalized $x(n)$ by the 10-pixel
spacing between electrodes, then measured the slope of $x(n)$ to
obtain $\alpha$. For this experiment $\alpha =$ 0.18.

The square width also increases linearly with the number of cycles
(data not shown); because we used a baseline threshold that
narrows the width of the distribution, however, the formula of
Eq.~\ref{e:width} no longer provides an accurate relationship
between $\sigma^2(n)$ and $\alpha$.

\begin{figure}[htbp]
\centering
\leavevmode
\psfig{file=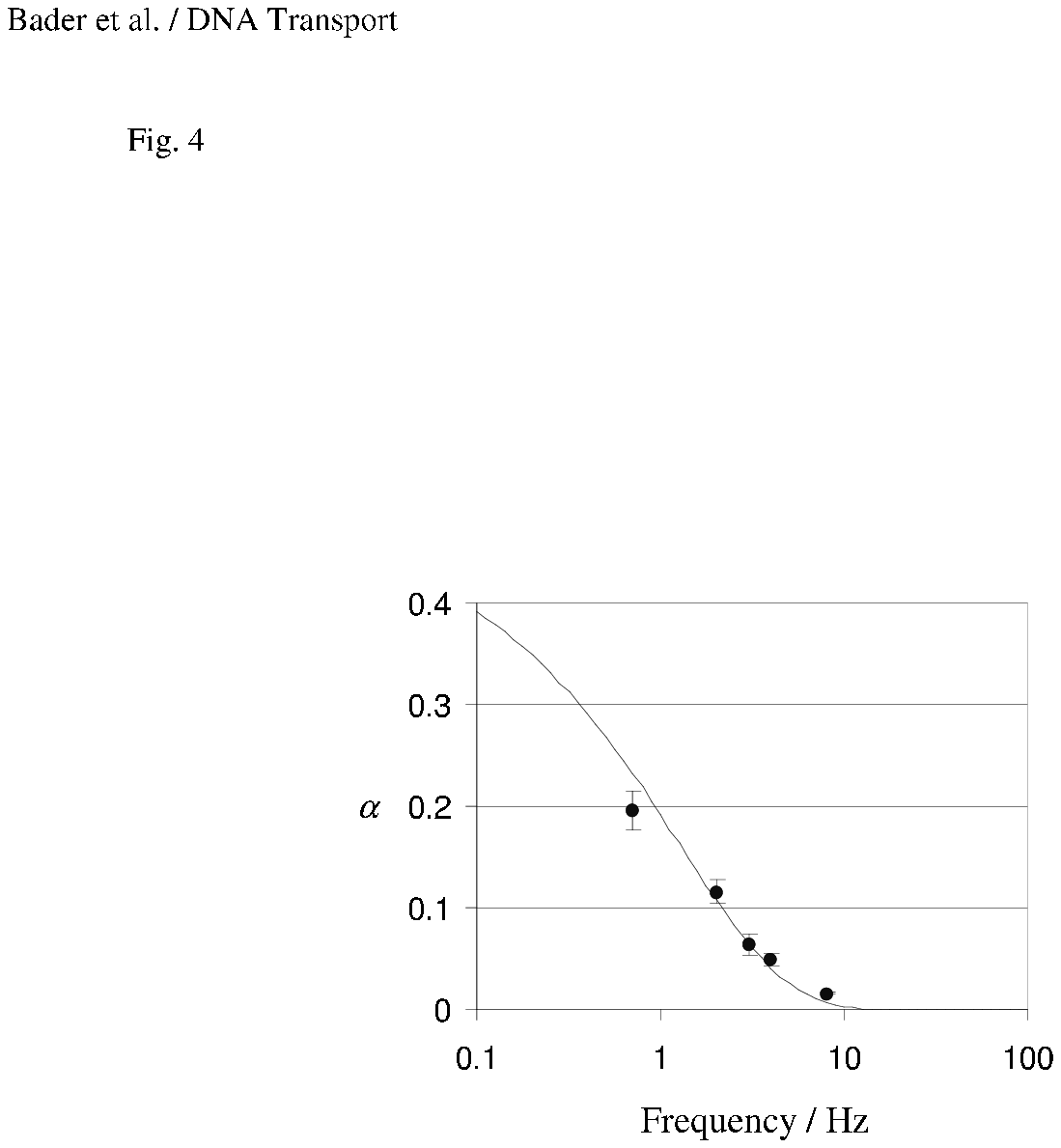,width=3in}
\caption{Experimental results for the transport of a DNA 50mer by
a device with 2~$\mu$m electrodes and 20~$\mu$m periodicity
(points with 1$\sigma$ error bars) are compared with theoretical
predictions (line).  The theory requires a single adjustable
parameter related to the diffusion constant of DNA.}
\label{f:experiment}
\end{figure}

Fig.~\ref{f:experiment} shows the experimental results for
$\alpha$, along with $\pm1\sigma$ error bars from repeated runs.
As predicted, $\alpha$ decreases with increasing frequency. Also
shown in Fig.~\ref{f:experiment} is a theoretical curve from
Eq.~\ref{e:alpha}.  Least-squares fitting to the data from all but
the highest frequency yielded $D(r/r_{\rm eff})^2 = 3.5\times
10^{-8}{\rm cm^2/sec}$.  The good agreement between the
experimental results and the single-parameter theoretical fit
supports our conclusion that transport is due to a Brownian
ratchet.

Furthermore, since $r_{\rm eff} \approx (2$--$3) r$, we find that
$D = 1.4$--$3\times 10^{-7}{\rm cm^2/sec}$.  This indirect measure
of the diffusion constant brackets the estimated diffusion
constant for a DNA 50mer close to a glass surface, $1.8 \times
10^{-7} {\rm cm}^2/{\rm sec}$ \cite{xy97}.

The experimental results demonstrating transport permit an
examination of the feasibility fabricating a device to provide
useful size-based separations of DNA.  Separating two chemical
species requires that they have different diffusion constants $D$
and $D'$ and different hopping probabilities $\alpha$ and
$\alpha'$. The resolution between the species, defined as
\begin{equation}
{\rm resolution} = \frac{| x(n) - x'(n) |} { 0.5 \cdot [\sigma(n)
+ \sigma'(n)] }, \label{e:resolution}
\end{equation}
improves as $n^{1/2}$.  The number of cycles required to reach the
resolution of 1 typical for DNA separation applications is
\begin{equation}
n = \frac{\alpha (1 - \alpha) }{(\alpha - \alpha')^2},
\label{e:n}
\end{equation}
where we have assumed that the two packets acquire a similar
width. The separation parameter $t_{\rm off}$ (and, through
$t_{\rm off}$, the quantities $\alpha$ and $\alpha'$) can be
selected to optimize various quantities associated with a resolved
separation, for example the device length, approximately $l \times
[\alpha/(\alpha - \alpha')]^2$, or the total separation time, $n
\times (t_{\rm on} + t_{\rm off})$.

A typical application requiring the separation of DNA fragments is
the analysis of single nucleotide polymorphisms (SNPs).  An SNP is
a position in the genome where multiple nucleotides are likely.
Characterizing genetic diversity through SNPs has applications
including the identification of genes for disease inheritance and
susceptibility, the development of personalized medicines, and the
documentation of human evolutionary history and migrations through
a genetic record
\cite{rm96,l96,cgc97,k97}.
Preliminary sets of thousands of SNPs, identified by
the 12~nt on either side of a polymorphism, have been reported
\cite{w98}.  Validating and detecting these SNPs can be
accomplished by resequencing specific 25~nt regions of the genome.

Here we investigate the use of a Brownian ratchet device for the
resequencing and detection application.  Calculations are based on
a device with feature size $r = 0.1\,\mu$m, periodicity 1~$\mu$m,
and a potential difference of $0.1~V$ between electrodes in the
on-state. We estimate the effective DNA diffusion constant
$D(r/r_{\rm eff})^2$ from the theoretical scaling for a
self-avoiding walk, $D \sim ({\rm length})^{-0.6}$
\cite{f53,d86}, and our experimental results for the 50mer.
The persistence length of single-stranded DNA is 4~nm, or 13.6~nt
\cite{persistencelength}, indicating that the self-avoiding walk
should be adequate (although not quantitative) for fragment sizes
we consider.

The detection of an SNP requires, at most, the ability to sequence
the 25~nt region surrounding the polymorphism, which can be
accomplished by separating a DNA 24mer from a 25mer respectively.
Using the theoretical scaling of diffusion constant with DNA
length, we extrapolate $D(r/r_{\rm eff})^2$ of $5.44 \times
10^{-8} {\rm cm}^2/{\rm sec}$ and $5.31 \times 10^{-8} {\rm
cm}^2/{\rm sec}$ for the 24mer and 25mer.  The time required to
focus the DNA at the start of each cycle is $t_{\rm on} \approx
1.8 \times 10^{-4} {\rm sec}$. The optimized separation
parameters, calculated using Eq.~\ref{e:n}, require $t_{\rm off} =
2.5 \times 10^{-4} {\rm sec}$ and 12,000 cycles for a total
separation time of 5.4~sec on a 1.25~cm chip.

In conclusion, we have fabricated a Brownian ratchet device that
is capable of transporting small DNA molecules in aqueous
solution, rather than the inconvenient gel and polymer solutions
required for electrophoresis.  This type of device could be used
as a pump component for transport or separation of charged species
in a microfabricated analysis chip.  Multiple miniaturized devices
can also be arrayed side-by-side for high-throughput operation.
Based on experimental measurements, we suggest the feasibility of
using this type of device for biological applications, for example
the validation of SNPs.

\acknowledgments

We wish to acknowledge the support of SBIR grant 1~R43~HG01535-01
from the National Human Genome Research Institute and Advanced
Technology Program award 1996-01-0141 from the National Institute
of Standards and Technology.  We acknowledge the assistance of
Rajen Raheja for image analysis.

\clearpage

\begin{references}
\bibitem[*]{jsb} Author to whom correspondence should be
addressed.
\end{references}
\end{document}